\documentclass[prl,showpacs,amsmath,preprint]{revtex4}

\usepackage{graphicx}
\usepackage[dvips]{color}
\usepackage{inputenc}

\begin{document}
\title{Quantitative Raman measurements of the evolution of the
    Cooper-pairs density with doping in Bi$_{2}$Sr$_{2}$CaCu$_{2}$O$_{8+\delta}$ superconductors}

\author{S. Blanc, Y. Gallais, A. Sacuto,  M. Cazayous and M. A. M\'easson}
\affiliation{Laboratoire Mat\'eriaux et Ph\'enom$\grave{e}$nes Quantiques (UMR 7162 CNRS),
Universit\'e Paris Diderot-Paris 7, Bat. Condorcet, 75205 Paris Cedex 13, France}

\author{G. D. Gu, J. S. Wen and Z. J. Xu}
\affiliation {Matter Physics and Materials Science, Brookhaven National Laboratory (BNL), Upton, NY 11973, USA}.

\date{\today}

\begin{abstract}

We report Raman measurements on $Bi_{2}Sr_{2}CaCu_{2}O_{8+\delta}$ single crystals which allow us to quantitavely evaluate the doping dependence of the density of Cooper pairs in the superconducting state. We show that the drastic loss of Cooper pairs in the antinodal region as the doping level is reduced, is concomitant with a deep alteration of the quasiparticles dynamic above $T_{c}$ and consistent with a pseudogap which competes with superconductivity. Our data also reveal that the overall density of Cooper pairs evolves with doping, distinctly from the superfluid density above the doping level $p_c=0.2$.

\end{abstract}


\maketitle

\par
\par

One of the most challenging issues in cuprate superconductors is to understand the low energy quasiparticles dynamics above and below the critical temperature $T_c$ as the Mott insulating state is approached by decreasing the doping level \cite {Anderson, OrensteinMillis}. On the overdoped side, below $T_c$, the $d-$wave superconducting gap develops with maximum values along the principal axes of the Brillouin zone, the antinodes and vanishes in the nodal regions, the diagonals of the Brillouin zone \cite{Damascelli}. On the underdoped side, above $T_c$ and below $T^{*}$, the pseudogap occurs in the antinodal regions \cite{Timusk_RPP99}. 
\par

Recent advances in angle resolved photoemission spectroscopy (ARPES) \cite{Lee, KondoPRL,KondoNature}, scanning tunneling spectroscopy (STS) \cite{Yasdani,Kohsaka}, $\mu-SR$ spectroscopy \cite{Khasanov} and Electronic Raman Scattering (ERS) \cite{LeTacon, Guyard_PRB, Guyard_PRL} have brought strong experimental evidences that superconductivity remains robust at the nodes even at low doping level while superconductivity is deeply altered at the antinodes. This manifests itself in ARPES, by the supression of the coherent spectral weight at the antinodes with underdoping \cite{KondoNature}. In STS, this is signalled by a shrinkage of the Bogoliubov quasiparticles arcs around the nodes \cite{Kohsaka} and in ERS, by the disappearance of the pair breaking peak in the antinodal Raman response as the doping level is reduced. \cite{LeTacon,Guyard_PRL,Opel,Gallais}. 
\par

These observations raise the question of the influence of doping on the k-space dependence of the superconducting properties and the relationship between the pseudogap and superconductivity. Our aim here, is to capture the doping evolution of the density of Cooper pairs in momentum space and to compare it with the superfluid density one. We also want to address whether a connection exists or not between the pseudogap and superconductivity.

\par

To achieve this goal, we have developed a careful and systematic experimental protocol which allows us to quantitatively compare the changes of the Raman spectra of $Bi_{2}Sr_{2}CaCu_{2}O_{8+\delta}$ ($Bi-2212$) compounds as a function of the doping levels. Using a simple relationship between the integrated superconducting Raman response and the density of Cooper pairs, we show that the density of Cooper pairs in the superconducting state strongly decreases with underdoping at the antinodes while it still sizeable in the nodal region even at low doping level. Simultaneously, our data reveal that in the normal state, the low energy quasiparticle dynamics is deeply altered as the doping is reduced in the antinodal region, whereas the nodal quasiparticles remain almost unaffected. 
\par
Below the doping level $p_c=0.2$, we find that the overall Cooper pairs density, $N_{Cp}$ decreases with underdoping in a similar way to the superfluid density deduced from $\mu-SR$ \cite{Uemura,Bernhard} magnetic penetration depth \cite{Panogopoulos2} and optical conductivity measurements \cite{Homes}. Above the doping level $p_c=0.2$, however, the two physical quantities appear to be disconnected.  

\par

The $Bi-2212$ single crystals were grown by using a floating zone method. The optimal doping sample with $T_c = 91~K$ was grown at a velocity of 0.2 mm per hour in air \cite{Wen_b}. In order to get the overdoped sample, the as-grown single crystal was put into a high pressured cell with $2000$ bars oxygen pressure and then was annealed at $350^{o}C$ to $500^{o}C$ for $72$ hours \cite{Mihaly}. In order to get underdoped sample, the optimal doping crystal was annealed at $350~^{o}C$ to $550~^{o}C$ for $72$ hours under vacuum of $1.3~10^{-6}~mbar$. The doping value $p$ is inferred from $T_c$ using Presland and Tallon's equation: $1-T_{c}/T_{c}^{max} = 82.6 (p-0.16)^{2}$ \cite{PreslandPhysicaC91}. $T_{c}$ has been determined from magnetization susceptibility measurements for each doping level.   

\par
Raman experiments have been carried out using a triple grating spectrometer (JY-T64000). The $B_{2g}$ and $B_{1g}$ geometries have been obtained from cross polarizations along the Cu-O bond directions and at 45$^o$ from them respectively \cite{sacuto_PRB00}. In these geometries we probe respectively, the nodal and antinodal regions of the Brillouin zone. All the measurements have been corrected for the Bose factor and the instrumental spectral response. 
\par
Special care has been devoted to make reliable quantitative comparisons between the Raman intensities of distinct crystals with different doping levels measured in the same geometry, and between measurements in distinct geometries for crystals with the same doping level. Obtaining intrinsic Raman measurements of crystals with various doping levels is a true challenge for experimentalists. It requires not only an extremely high level of control of the crystal surface quality, the optical set up but also the knowledge of the optical constants for each crystal studied. 
\par
In order to overcome these difficulties, we have first chosen to work on $Bi-2212$ system rather than on the $Hg-1201$ ($HgBa_{2}CuO_{4+\delta}$) one as previously \cite{LeTacon,Guyard_PRB} because $Bi-2212$ crystals can be easily cleaved providing large homogeneous surfaces ($\approx mm^{2}$). We have performed all the measurements during the same run and the crystals with various doping levels have been mounted on the same sample holder in order to keep the same optical configuration. With a laser spot of $\approx 50\mu m$ diameter, we have measured Raman intensity variations of less than $5\%$ from one point to another on the same cleaved surface. Crucially, we have also observed only weak intensity changes for two distinct crystals of the same nominal doping level mounted side by side on the sample holder of the cryostat. These observations give us confidence that the doping dependence of the Raman intensity variations reported here are intrinsic. Finally, the Raman cross-section at each doping level was obtained by correcting the Raman response function for the optical constants, using the following expression for the correction factor \cite{Reznik}:
$\frac{[\alpha(\omega_{s})+\alpha(\omega_{i})].n(\omega_{s})^2}{T(\omega_{s}).T(\omega_{i})}$ where $\alpha$ refers to the absorption, $T$ to the transmission at the air-sample interface and $n$ to the complex refractive index whose components have been determined from spectroscopic ellipsometry measurements.  
\par

\begin{figure}
\begin{center}
\includegraphics[width=8cm]{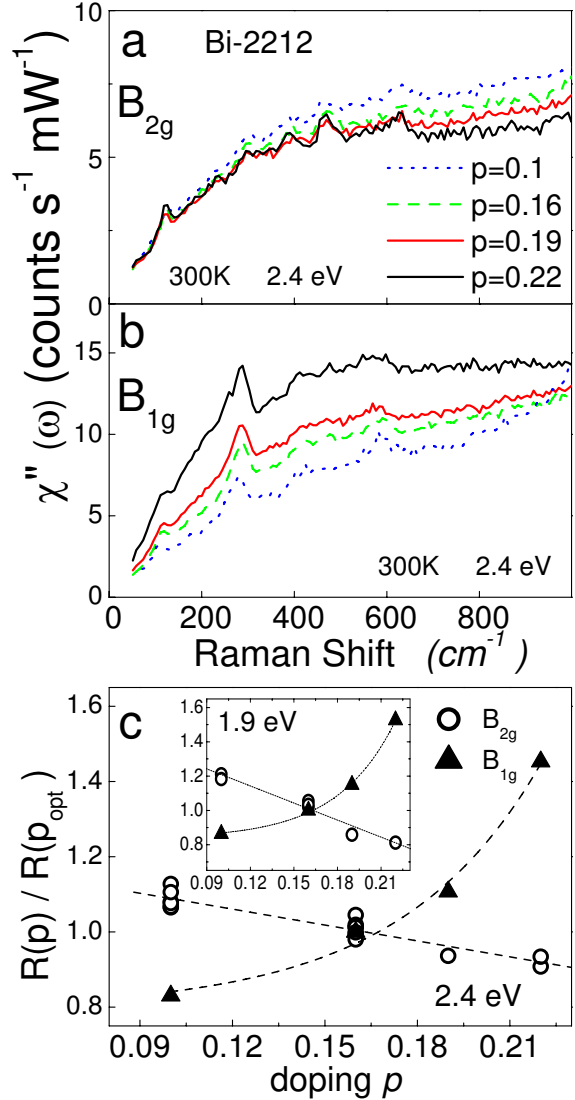}
\end{center}\vspace{-8mm}
\caption{(Color online) 
Intrinsic Raman response functions of $Bi-2212$ single crystals in (a) $B_{2g}$ (nodal) and (b) $B_{1g}$(antinodal) geometries with various doping levels $p$. Each spectrum has been obtained from an average over 10 spectra measured on different regions of the sample surface at $300~K$. (c) $R(p)=\int^{1000 ~cm^{-1}}_{50~cm^{-1}}{\chi_{\mu}^{''}(\omega,p) d\omega}$ where $\mu$ denoted the $B_{1g}$ or $B_{2g}$ geometries. Integrals $R(p)$ is normalized to the optimal one $R(p_{opt})$.}

\label{fig1}
\end{figure}
 
We first focus on the evolution of the quasiparticle dynamics with doping before studying the superconducting state. In Figure 1 (a) and (b) we display the normal state Raman responses of $Bi-2212$ single crystals for four doping levels.  Figures 1 (a) and (b) show that the $B_{2g}$ electronic background intensity is almost the same as a function of doping level up to $400~cm^{-1}$ while the $B_{1g}$ electronic continuum intensity is drastically reduced as the doping level decreases \cite{note1}. Correspondingly the low energy $B_{2g}$ slope varies only weakly with doping while the low energy $B_{1g}$ slope strongly decreases with underdoping. More surprisingly, the $B_{2g}$ electronic background intensity slightly increases as the doping level decreases while the $B_{1g}$ electronic continuum is drastically reduced. This phenomenon is also observed by using different excitation lines ($2.56$ and $1.93~eV$) (see Fig.1 (c)) which eliminates the possibility of a resonant Raman effect.
 \par
In order to quantify these observations, in Fig.1 (c) we plot the integrals , $R(p)$, of the $B_{2g}$ and $B_{1g}$ Raman response functions normalized to the one at the optimal doping. We observe a strong decrease of the $B_{1g}$ response (about $65 \%$) with underdoping while the $B_{2g}$ response exhibits a smaller change in the opposite way (about $20 \%$) from $p= 0.22$ to $0.1$.  
\par
Assuming a Drude-like Raman response function in the framework of Landau theory of interacting particles, the Raman response in the normal state leads to \cite{Devereaux}: $N_F\frac{(Z\Lambda)_k^2\Gamma_k\omega}{\Gamma_k^{2}+\omega^2}$ where $N_F$ is the density of state at the Fermi level, $\omega$ the Raman shift, $\Gamma_k$ the quasiparticles scattering rate and $(Z\Lambda)_k$ the renormalized quasiparticle spectral weight where $\Lambda$ takes into account final state interactions. The low energy slope is then proportionnal to $\frac{(Z\Lambda)_k^2}{\Gamma_k}$ ratio which indicates that the lowering of the quasiparticles spectral weight and/or the enhancement of the scattering rate at the antinodes are responsible for the strong decrease of the low energy $B_{1g}$ slope with underdoping. In sharp contrast, the nodal quasiparticles are mostly unaffected by the doping. Combined with earlier and recent ARPES data \cite{Shen_Science05, KondoNature,Campuzano_com} and previous Raman measurements \cite{Opel,Venturini,Sugai,Gallais} this shows that the quasiparticle spectral weight in the nodal region is weakly doping dependent.

\par
 
In Figure 2 we display the $B_{2g}$ and $B_{1g}$ superconducting responses of $Bi-2212$ for several doping levels in the superconducting and normal states. At low energy, the slope of the $B_{2g}$ superconducting response is almost doping independent as suspected previously in $Hg-1201$ system \cite{LeTacon}. 

At higher energy, we focus on the $B_{1g}$ and $B_{2g}$ superconducting peak areas deduced from the substraction between the superconducting and the normal Raman responses  and displayed in grey in Fig. 2(a) and (b). 
 Our data reveal a strong decrease of the $B_{1g}$ superconducting peak area with underdoping. It disappears close to $p= 0.1$ while the $B_{2g}$ superconducting peak area slightly increases from $p=0.22$ to $0.19$ and then remains almost constant as the doping level is reduced down to $0.1$. On the overdoped side (above $p=0.16$), the $B_{1g}$ superconducting peak area is predominant with respect to the $B_{2g}$ one while this is the opposite on the underdoped side.

 \par
What is the meaning of the superconducting peak area? For a non interacting Fermi liquid in the framework of BCS theory, the Raman response in the superconducting state is given by \cite{Klein,Devereaux}: 

              $$\chi^{,,}_{\mu}(\omega) =\pi  \sum_{k}(\gamma^{\mu}_{k})^{2}\tanh (\frac{E_{k}}{2k_{B}T})\frac{\left|\Delta_k\right|^{2}}{E_{k}^{2}}\delta(\omega-2E_{k})$$
              
where $ \mu $ refers to the $B_{1g}$ and $B_{2g}$ geometries, $ \gamma^{\mu}_{k}$ is the Raman vertex, $ \Delta_k$, the superconducting gap, $E_{k}$, the quasiparticle energy and $k_B$, the Boltzman constant. It is then straightforward to show that the integral of the Raman response over $\omega$ when $T$ tends to zero, gives: 
             
               $$\int{\chi^{,,}_{\mu}(\omega)d\omega}=\pi  \sum_{k}(\gamma^{\mu}_{k})^{2}\frac{\left|\Delta_k\right|^{2}}{E_{k}^{2}}$$
          
The sum $\sum_{k}\frac{\left|\Delta_k\right|^{2}}{E_{k}^{2}}$ is equal to $4\sum_{k}(u_kv_k)^2$ where $v_k^2$ and $u_k^2$ are the probabilities of the pair $(k\uparrow,-k\downarrow)$ being occupied and unoccupied respectively. This sum is non-vanishing only around the Fermi energy $E_F$ in the range of $2\Delta_k$ \cite{deGennes}. This quantity corresponds to the density of Cooper pairs, formed around the Fermi level as the gap is opening \cite{Leggett}. A priori, the density of Cooper pairs is distinct from the superfluid density which is just the total carrier density at $T=0~K$. The integral of the Raman response is then proportional to the density of Cooper pairs, weighted by the square of the Raman vertex which selects specific area of the Brillouin zone: the nodal or the antinodal regions. 

\par

\begin{figure}
\begin{center}
\includegraphics[width=8cm]{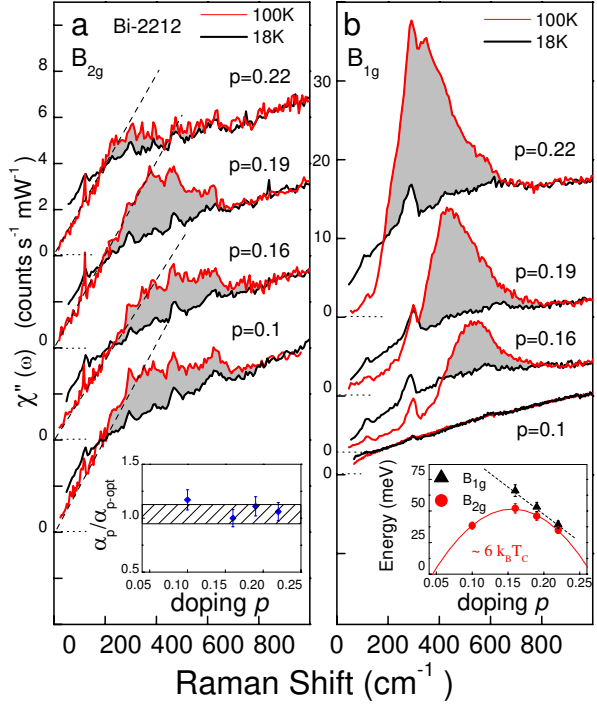}
\end{center}\vspace{-8mm}
\caption{ (Color online)
 Raman response functions in the normal and superconducting states in (a) $B_{2g}$ and (b) $B_{1g}$ geometries for distinct doping levels. The grey zones correspond to the subtraction between the superconducting and the normal Raman responses under the superconducting pair breaking peaks. The insets exhibit the doping evolution of the normalized $B_{2g}$ low energy slope $\alpha_{p}/\alpha_{p-opt}$ and the doping dependences of the $B_{1g}$ and $B_{2g}$ peak energies extracted from asymetrical gaussian fits. The slope of the nodal $B_{2g}$ superconducting Raman susceptibility, $\alpha_{p}$, is only weakly doping dependent and its variation is less than $20\%$ between $p= 0.1$ and $p= 0.22$.}
\label{fig2}
\end{figure}

Applying this analysis to our data reveals that the superconducting peak area (in grey in Fig.2) provides a direct estimate of the density of Cooper pairs in the nodal and antinodal regions \cite{note2}. The data reported in Fig. 3 (a) show that the density of Cooper pairs is strongly anisotropic in the $k-$space as a function of doping level. At low doping level, the density of Cooper pairs becomes very small at the antinodes and vanishes below $p=0.1$, while it is still sizable around the nodes. Therefore we are led to conclude that Cooper pairs are k-space localized in the nodal region at low doping level forming k-space Cooper pairs islands on the underdoped regime of cuprates. This is consistent with the picture where most of the supercurrent is carried out by electrons' small patches centered on the nodal points on the underdoped regime as proposed by Ioffe and Millis \cite{Ioffe_98}. 
\par
Further, comparison between Figs.1 and 2 reveals that the loss of Cooper pairs density at the antinodes below $T_c$ is concomitant with the strong alteration of the quasiparticles dynamics at the antinodes above $T_c$ where the pseudogap develops as the doping level diminishes. If the pseudogap state was due to preformed pairs above $T_c$, these preformed pairs (if they are involved in the superconducting condensate below $T_c$) should give a sizeable density of Cooper pairs in the antinodal region and an intense superconducting peak in contradiction to our findings. As a consequence, our data support the view that the pseudogap is harmful to the Cooper pairs formation and acts as a "`foe"' of high $T_c$ superconductivtiy \cite{Norman_adv_05}. This view is consistent with recent ARPES data \cite{KondoNature} which reveal a direct correlation between the opening of the pseudogap and the decrease of the spectral weight of the superconducting coherent peak.

\par  
\begin{figure}
\begin{center} 
\includegraphics[width=8cm]{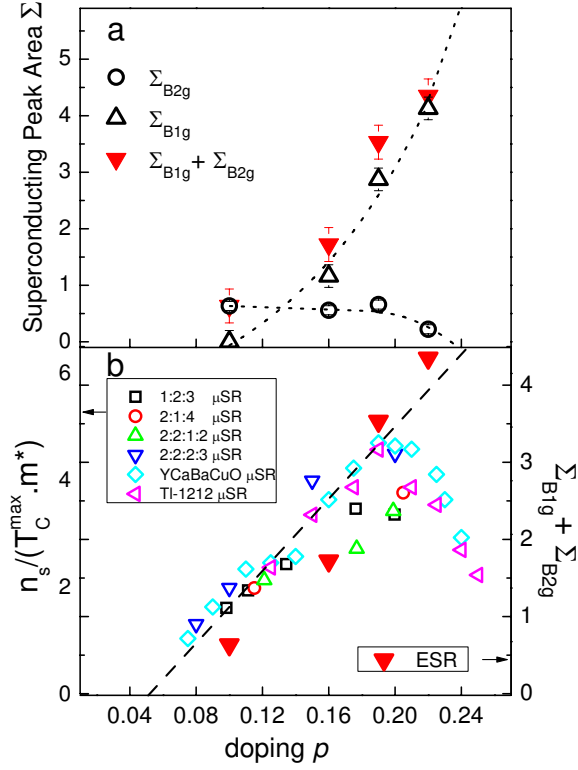}
\end{center}\vspace{-5mm}
\caption{(Color online)
 (a) Doping dependences of the $B_{1g}$ and $B_{2g}$ superconducting peaks areas ($\Sigma_{B_{1g}}$,$\Sigma_{B_{2g}}$) and their sum $\Sigma_{B_{1g}+B_{2g}}$ (in $Cts/(s.mW).cm^{-1}$) deduced from the subtraction between the superconducting and the normal Raman reponses. (b) Doping dependences of  $\Sigma_{B_{1g}+B_{2g}}$ and the normalized superfluid density ($\frac{n_{s}}{T_{c}^{max}.m*}$) deduced from \protect\cite{Uemura}. We have used the Presland's relation \protect\cite{PreslandPhysicaC91} to convert $T_{c}$ as a function of doping in the Uemura's plot and finally divided $\frac{n_{s}}{m*}$ by $T_{c}^{max}$ in order to compare to different cuprate families.} 
\end{figure}

\label{fig3}

If we now concentrate on the doping dependence of the sum of the superconducting peak areas ($\Sigma_{B_{1g}}+\Sigma_{B_{2g}}$) shown in Fig. 3 (a), we note that the sum $\Sigma_{B_{1g}}+\Sigma_{B_{2g}}$ decreases with underdoping suggesting a decrease of the \textit {overall} Cooper pairs density, $N_{Cp}$ with doping.
\par
To further substantiate this point, we have replotted in Fig. 3(b) the superfluid density $\frac{n_{s}}{m*}$ normalized to $T_{c}^{max}$ obtained from ${\mu}-SR$ \cite{Uemura,Bernhard} as a function of doping level, $p$, instead of $T_c$ as it is usually reported. We observe that for each cuprates family considered here ($Y-123$, $La-124$, $Bi-2212$,$Tl-1212$ and $Bi-2223$), this ratio increases approximatively linearly with doping up to $p_c=0.2$ while beyond $p_c$ it decreases. Comparing with our estimations of the overall Cooper pairs density $N_{Cp}$, it appears that $N_{Cp}$ behaves like the superfluid density ($N_{Cp}\propto p$) up to $p_c$ but, by contrast continues to increase above $p_c$. The decrease of $N_{Cp}$ and $\frac{n_{s}}{m*T_{c}^{max}}$ below $p_c$ is consistent with a scenario where the pseudogap develops below $p_c$ and suppresses both superfluid density and the density of Cooper pairs at the antinodes. 
\par

In summary, we have developed a reproducible experimental protocol which allows us to get reliable comparison between the Raman spectral intensities of several crystals with distinct doping levels in both the $B_{2g}$ (nodal) and $B_{1g}$ (antinodal) geometries. From the integrated superconducting response, we can then quantitatively evaluate, the density of Cooper pairs nearby the Fermi level and tracks its doping evolution in the nodal and antinodal regions. Moreover, we find that the loss of Cooper pairs density below $T_c$, at the antinodes, is concomitant with a deep alteration of the quasiparticle dynamics above $T_c$ as the doping level is reduced. This strongly suggests that the pseudogap state competes with the superconducting state. 

\par

\section*{Acknowledgements}

We are grateful to A. Georges, J. C. Campuzano, A. Millis, C. Ciuti, G. Blumberg, Y. Sidis and P. Monod for very helpful discussions. We acknowledge support from the ANR grant No. BLANC07-1-183876 GAPSUPRA. Correspondences and requests for materials should be adressed to A.S (alain.sacuto@univ-paris-diderot.fr).


\begin{thebibliography}{9}

\bibitem{Anderson} P. W. Anderson, Science 235, 1196 (1987).

\bibitem{OrensteinMillis} J. Orenstein and A.J.Millis, Science \textbf{288}, 468 (2000).

\bibitem{Damascelli} A.Damascelli, Z.Hussain and Z.X. Shen, Rev. Mod. Phys., \textbf{75}, 473 (2003). 

\bibitem{Timusk_RPP99} T. Timusk and B. W. Statt, Rep. Prog. Phys., \textbf{62}, 61-122 (1999).

\bibitem{Lee} W. S. Lee, I. M. Vishik, K. Tanaka, D. H. Lu, T. Sasagawa, N. Nagaosa, T. P. Devereaux, Z. Hussain and  Z.-X. Shen, Nature \textbf{450}, 81 (2007).

\bibitem{KondoPRL}T. Kondo, T. Takeuchi, A. Kaminski, S. Tsuda and S. Shin, Phys.Rev. Lett \textbf{98}, 267004 (2007).

\bibitem {KondoNature} T. Kondo, R. Khasanov, T. Takeuchi,J. Schmalian and A. Kaminski, Nature \textbf{457}, 296 (2009).

\bibitem{Yasdani} A.Pushp, C.V. Parker, A. N. Pasupathy, K. K. Gomes, S. Ono, J. Wen, Z.J. Xu, G.D. Gu and A. Yazdani , Science \textbf{324}, 1689 (2009)

\bibitem {Kohsaka} Y. Kohsaka, C. Taylor, P. Wahl, A. Schmidt, Jhinhwan Lee, K. Fujita, J. W. Alldredge, K. McElroy, J. Lee, H. Eisaki, S. Uchida, D.H. Lee and  J. C. Davis Nature \textbf{454}, 1072 (2008).

\bibitem {Khasanov} R. Khasanov, T. Kondo, S. Strässle, D. O. Heron, A. Kaminski, H. Keller, S. L. Lee and T. Takeuchi, Phys Rev. Lett. \textbf{101}, 227002 (2008).

\bibitem{LeTacon}  M. Le Tacon, A. Sacuto, A. Georges, G. Kotliar, Y. Gallais, D. Colson and A. Forget, Nat. Phys. \textbf{2}, 537 (2006).

\bibitem{Guyard_PRB}  W. Guyard, M. Le Tacon, M. Cazayous, A. Sacuto, A. Georges, D. Colson and A. Forget, Phys. Rev. B \textbf{77}, 024524 (2008).

\bibitem{Guyard_PRL}  W. Guyard, A. Sacuto, M. Cazayous, Y. Gallais, M. Le Tacon, D. Colson and A. Forget 

\bibitem{Uemura}  Y. J. Uemura, {\it et al.}, Phys. Rev. Lett.\textbf{62}, 2317 (1989).

\bibitem{Panogopoulos2} C. Panagopoulos, J. L. Tallon, B. D. Rainford, T. Xiang, J. R. Cooper and C. A. Scott, Phys. Rev. B \textbf{66}, 064501, (2002). 

\bibitem{Bernhard} C. Bernhard, J. L. Tallon, Th. Blasius, A. Golnik and Ch. Niedermayer, Phys. Rev. Lett. \textbf{86}, 1614, (2001).

\bibitem{Homes} C. C. Homes, S. V. Dordevic, M. Strongin, D. A. Bonn, R. Liang, W. N. Hardy, S. Komiya, Y. Ando, G. Yu, N. Kaneko, X. Zhao, M. Greven, D. N. Basov and T. Timusk, Nature \textbf{430}, 539 (2004). 

\bibitem{Wen_b} J. S. Wen,Z.J. Xu, G.Y. Xu, M. Hückera, J.M. Tranquada and G.D. Gu, J. of Crystal Growth. \textbf{310}, 1401 (2008).

\bibitem{Mihaly}  L. Mihaly, C. Kendziora, J. Hartge, D. Mandrus and L. Forro, Rev. Sci. Instrum. \textbf{64}, 2397 (1993).

\bibitem{PreslandPhysicaC91} M. R. Presland, J. L. Tallon, R. G. Buckley, R. S. Liu and N. E. Flower, Physica C \textbf{176}, 95 (1991).

\bibitem{note1} Excepted for $p=0.1$ above $900~cm^{-1}$ where the double magnon contribution appears due to the AF Mott phase proximity.

\bibitem{sacuto_PRB00} A. Sacuto, J. Cayssol, D. Colson and P. Monod, Phys. Rev. B \textbf{61}, 7122 (2000).

\bibitem{Reznik}  D. Reznik, S. L. Cooper, M. V. Klein, W. C. Lee,D. M. Ginsberg,A. A. Maksimov, A. V. Puchkov,I. I. Tartakovskii and S-W. Cheong, Phys. Rev. B \textbf{48}, 7624 (1993).

\bibitem{Devereaux} T. P. Devereaux and R. Hackl, Rev. Mod. Phys. \textbf{79}, 175 (2007).

\bibitem{Shen_Science05} K. M. Shen,F. Ronning, D. H. Lu, F. Baumberger, N. J. C. Ingle, W. S. Lee, W. Meevasana, Y. Kohsaka, M. Azuma, M. Takano, H. Takagi and Z.-X. Shen, Science \textbf{307}, 901-904 (2005).

\bibitem{Campuzano_com} J. C. Campuzano private communication, august (2008).

\bibitem{Opel} M. Opel, R. Nemetschek, C. Hoffmann, R. Philipp, P. F. Müller, R. Hackl, I. Tütto, A. Erb, B. Revaz, E. Walker, H. Berger and L. Forró Phys. Rev. B \textbf{61}, 9752-9844 (2000).

\bibitem{Venturini}  F. Venturini, M. Opel, T. P. Devereaux, J. K. Freericks, I. Tütto, B. Revaz, E. Walker, H. Berger, L. Forró and R. Hackl ,Phys. Rev. Lett.  \textbf{89}, 107003 (2002).

\bibitem{Sugai} S. Sugai, H. Suzuki, Y. Takayanagi, T. Hosokawa and N. Hayamizu, Phys. Rev. B \textbf{68}, 184504 (2003).

\bibitem{Gallais} Y. Gallais, A. Sacuto, T. P. Devereaux and D. Colson, Phys. Rev. B \textbf{71}, 012506 (2005).

\bibitem{Klein} M. V. Klein  and  S. B. Dierker,  Phys. Rev. B \textbf{29}, 4976 (1984).

\bibitem{deGennes}  P. G. de Gennes,  Superconductivity of Metals and Alloys, Addison Wesley Publishing Company, (1966).

\bibitem{Leggett} A. J. Legett,  Quantum Liquids, Bose Condensation and Cooper Pairing in Condensed-Matter Systems, Oxford University Press, (2006).

\bibitem{note2} We can note that if we had considered the complete integration of the superconducting responses we shoud have detemined a cut off at high energy level in order to fix the integral limits and we will obtain nearly the same doping evolution for the superconducting peak areas shown in Fig.3.

\bibitem{Ioffe_98} L. B. Ioffe and A. J. Millis, Journal of Phys. and Chem. of Solids \textbf{63}, 2259, (2002).

\bibitem{Norman_adv_05} M. R. Norman, D. Pines and C. Kallin, Adv. Phys. \textbf{54}, 715, (2005).

\end{thebibliography}
\end{document}